\documentclass[twocolumn,aps,amssymb,,superscriptaddress]{revtex4}
\usepackage{graphicx}

\begin{document}
\title{Coupling of phonons and electromagnons in GdMnO$_3$
}
\author{A. Pimenov}
\author{T. Rudolf}
\author{F. Mayr}
\author{ A. Loidl}
\affiliation{Experimentalphysik V, EKM, University of Augsburg,
86135 Augsburg, Germany} %
\author{A.$\>$A. Mukhin}
\affiliation{General Physics Institute of the Russian Acad. of
Sciences, 119991 Moscow, Russia}%
\affiliation{Experimentalphysik V, EKM, University of Augsburg,
86135 Augsburg, Germany}
\author{A. M. Balbashov}
 \affiliation{Moscow Power Engineering Institute, 105835 Moscow, Russia}
\date{\today}

\begin{abstract}
The infrared and Terahertz properties of GdMnO$_3$ have been
investigated as function of temperature and magnetic field, with
special emphasis on the phase boundary between the incommensurate
and the canted antiferromagnetic structures. The heterogeneous
incommensurate phase reveals strong magnetodielectric effects,
characterized by significant magnetoelectric contributions to the
static dielectric permittivity and by the existence of
electrically excited magnons (electromagnons). In the commensurate
canted antiferromagnetic phase the magnetoelectric contributions
to the dielectric constant and electromagnons are suppressed. The
corresponding spectral weight is transferred to the lowest lattice
vibration demonstrating the strong coupling of phonons with
electromagnons.
\end{abstract}

\pacs{75.80.+q,75.47.Lx,78.30.-j,75.30.Ds}

\maketitle

Multiferroic materials with the simultaneous occurrence of
magnetism and ferroelectricity, are a hot topic in recent
solid-state research. They provide interesting and spectacular
physical properties and promise attractive applications
\cite{hill,fiebig,khomskii}. Multiferroic behavior occurs in a
variety of systems originating from very different physical
mechanisms, including materials with independent magnetic and
ferroelectric subsystems, like some boracites, Aurivillius phases,
hexagonal manganites, and the lone-pair ferroelectrics with
magnetic ions \cite{khomskii}. Recently a new class of
multiferroics, namely ferromagnetic sulfospinels with relaxor-like
short range ferroelectric (FE) order have been detected
\cite{peter} with a strong coupling of the electric and magnetic
properties al low frequencies. In these spinel compounds the
ferromagnetism is induced via strong indirect exchange
interaction, but the origin of ferroelectricity remains unclear
sofar. Finally, in the perovskite manganites there is robust
experimental evidence \cite{kimura,kimura05} that the onset of
helical magnetic order induces spontaneous FE polarization
\cite{kenzelmann,arima}. Dzyaloshinskii-Moriya type interactions
have been utilized to explain the ferroelectricity which is
induced by the helical spin structure \cite{katsura05,mostovoy,
sergienko}. A similar spin-driven ferroelectricity is believed to
be operative in Ni$_3$V$_2$O$_8$ \cite{lawes}.

After having established the ground-state properties of this
interesting class of materials, the study of their novel dynamic
properties will significantly enhance our knowledge of the
magneto-electric (ME) coupling \cite{katsura}. Magnons are the
characteristic excitations of magnetic structures, while soft
phonons as inferred by the Lyddane-Sachs-Teller relation condense
at canonical ferroelectric phase transitions. It seems clear that
soft phonons cannot be relevant excitations in the ferroelectric
manganites, as (improper) ferroelectricity is induced by the
magnetic order coupled to the lattice. Recently it has been shown
that electro-magnons are relevant collective modes in this new
class of ferroelectrics \cite{nphys}. Electromagnons are spin
waves that are excited by an ac electric field. By measurements in
TbMnO$_3$ and GdMnO$_3$ it has been documented that these new
excitations exist not only in the magnetic phase characterized by
the helical spin structure, but also in the longitudinally
modulated (sinusoidal) structure, provided that a "helical-type"
vector component of the spin-wave is dynamically induced via the
ac electric field \cite{nphys}.

In this Letter we present detailed investigations of the Terahertz
and FIR properties of GdMnO$_3$. We investigate electromagnons and
phonons as function of temperature and magnetic field. We provide
striking experimental evidence that i) electromagnons are strongly
coupled to phonons with a considerable shift of optical weight
between these excitations and ii) electromagnons contribute
considerably to the static dielectric constant.

Single crystals of GdMnO$_3$ have been prepared using the
floating-zone method with radiation heating. The samples have been
characterized using X-ray, magnetic and dielectric measurement
\cite{jo}. The basic properties of our samples agree well with the
results obtained by other groups \cite{goto,kimura05}. The
experiments at Terahertz frequencies were carried out in a
Mach-Zehnder interferometer \cite{volkov} which allows the
measurements of transmittance and phase shift. The absolute values
of the complex dielectric permittivity
$\varepsilon^*=\varepsilon_1+i\varepsilon_2$ were determined
directly from the measured spectra using the Fresnel optical
formulas for the complex transmission coefficient. The spectra in
the infrared frequency range have been obtained using a Bruker
IFS-113 Fourier-transform spectrometer. The experiments in
external magnetic fields were performed in a superconducting
split-coil magnet with polyethylene windows allowing to carry out
reflectance experiments in magnetic fields up to 7\,T.

\begin{figure}[]
\includegraphics[width=7.5cm,clip]{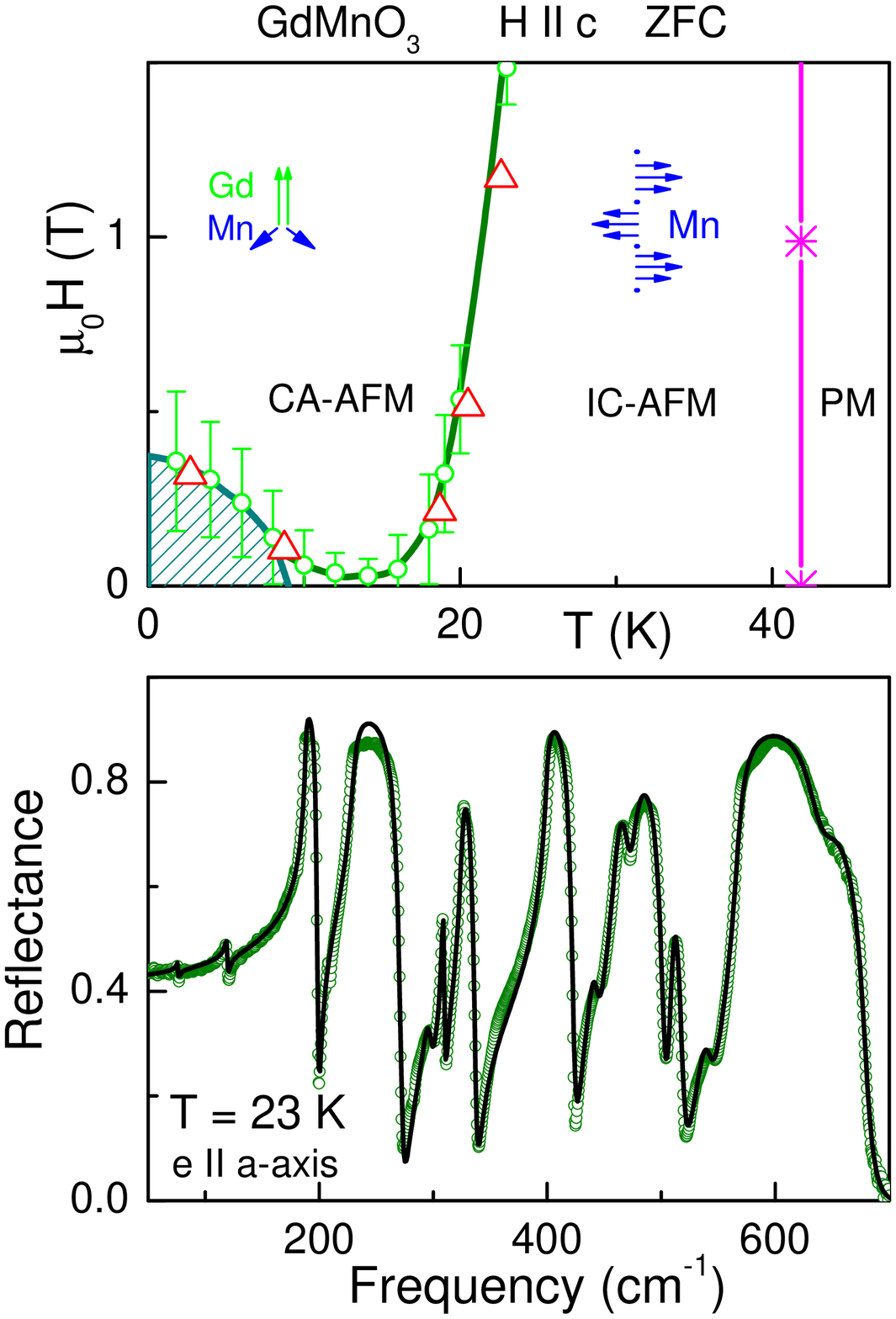}
\caption{(color online) Upper panel: Magnetic phase diagram of
GdMnO$_3$ for $H\|$c for zero-field cooling (ZFC). Circles and
stars: results of magnetization measurements, triangles: data from
dielectric experiments \cite{nphys}. Lines are drawn to guide the
eye. PM - paramagnetic, IC-AFM - incommensurate antiferromagnetic
(probably sinusoidally modulated \cite{kimura05}), CA-AFM - canted
antiferromagnetic structure. The
hatched region indicates the ordering of Gd-sublattice. \\
Lower panel: Reflectance of GdMnO$_3$ with polarization along the
a-axis and at phonon frequencies. Symbols: experiment, solid line:
model calculation assuming the sum of 15 Lorentzians (see text).}
\label{fphase}
\end{figure}

The upper panel of Fig. \ref{fphase} shows the H-T phase diagram
of GdMnO$_3$ for H $\|$c in the zero-field-cooled (ZFC) regime
\cite{nphys} which basically coincide with the diagrams published
previously \cite{jo,kimura05}. GdMnO$_3$ is paramagnetic above
$T_{\rm N}\simeq 42\ $K and transforms into an incommensurate
antiferromagnetic (IC-AFM) state below this temperature. Depending
upon the value of the external magnetic field along the c-axis,
the magnetic state becomes canted antiferromagnetic (CA-AFM) below
20 K. For $T < 9$ K and in low fields the spin structure reveals
increasing complexity due to an additional ordering of
Gd-subsystem. This region, which is indicated by a hatched area in
the phase diagram of the upper panel of Fig. \ref{fphase}, is not
further discussed in the course of this work. There is one
important difference compared to the $(H,T)$ phase diagram
published in Ref. \cite{jo}: under zero-field cooling conditions
no phase transition occurs and the AC-AFM state remains stable
down to 9\ K. This fact allows to switch between the
magnetoelectric IC-AFM and the CA-AFM phases with very low fields.
The IC-AFM region is especially interesting from the spectroscopic
point of view, because unusual excitations of mixed
magnetoelectric origin (electromagnons) exist \cite{nphys}. As
have been shown previously, the electromagnons are magnons which
can be excited by the electric component of the electromagnetic
wave. These excitations are suppressed in the CA-AFM state. It is
the aim of this work to study magnetic-field and temperature
dependence of electromagnons in a broad frequency range and to
investigate their possible coupling to phonon modes.

The lower panel of Fig. \ref{fphase} shows the reflectance
spectrum of GdMnO$_3$ at phonon frequencies and for the ac
electric field component $\tilde{e}$ parallel to the
crystallographic a-axis. We note that this direction of the
electric field reveals large effects in the temperature and field
dependence of the dielectric constant \cite{kimura05} and is
especially important for the magnetodielectric effects in this
compound. The solid line in the lower panel of Fig. \ref{fphase}
has been calculated using a sum of 15 Lorentzians. Here ten
strongest excitations correspond to phonons polarized along the
a-axis at frequencies 119, 188, 231, 308, 325, 400, 460, 475, 509,
and 568 cm$^{-1}$. Weaker features at 296, 441, 539, and 641
cm$^{-1}$ are due to leakage of the b-axis component, most
probably because of a small misorientation of the crystal. The
weak excitation at 76 cm$^{-1}$ is of unknown origin and might be
due to a forbidden exchange excitations of Mn or Gd subsystem,
which becomes IR allowed via strong magnetoelastic coupling.

\begin{figure}[]
\includegraphics[width=7.5cm,clip]{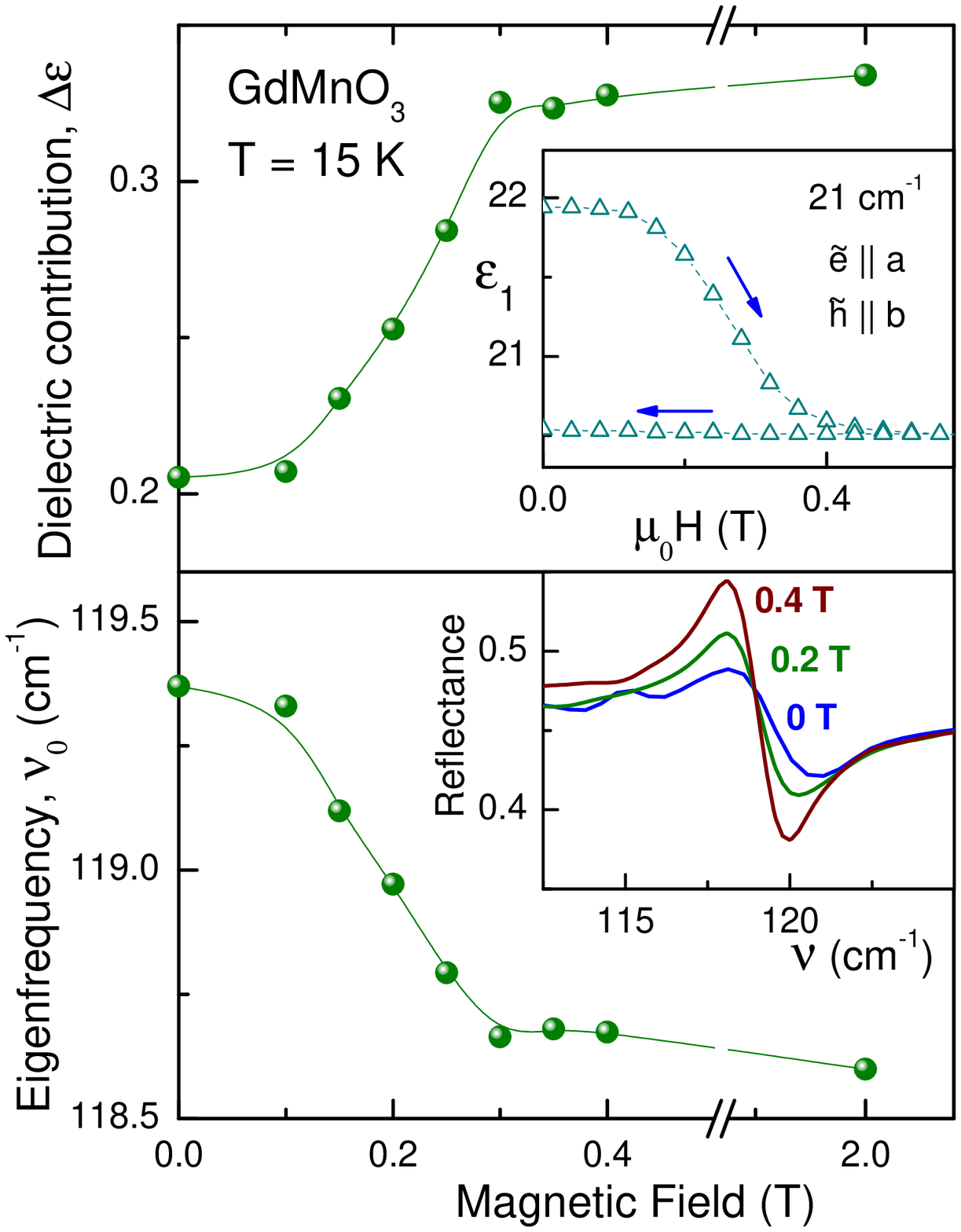}
\caption{(color online) Magnetic field dependence of the
parameters of the phonon at 119 cm$^{-1}$ crossing the boundary
between IC-AFM and CA-AFM phases. Upper panel: dielectric
contribution ($\Delta \varepsilon$), lower panel: eigenfrequency
($\nu_0$). The lower inset shows the reflectance spectra at
typical magnetic fields. The upper inset shows the changes in
dielectric permittivity at a frequency close to electromagnon,
which demonstrates direct connection of both effects.}
\label{fpar}
\end{figure}

Except for the lowest frequency phonon at 119 cm$^{-1}$, no
measurable changes in the phonon parameters have been detected
between different magnetic phases. On the contrary, the 119
cm$^{-1}$ phonon, which is observed for a-axis polarization only,
reveals substantial changes between  the IC-AFM and CA-AFM phases.
We recall that the peculiarity of this transition into the CA-AFM
phase is the occurrence of strong magnetodielectric effects, e.g.
the magnetic field dependence of the a-axis dielectric constant in
the frequency range from zero to about one Terahertz ($\sim 40$
cm$^{-1}$) \cite{nphys}. As have been shown previously
\cite{nphys}, these effects can be directly related to the
existence of electromagnons at $\nu \sim 20$ cm$^{-1}$.

The magnetic field dependence of the 119 cm$^{-1}$ phonon
excitation on crossing the IC-CA magnetic phase boundary is shown
in the lower inset of Fig. \ref{fpar} which represents the
reflectance spectra at three typical fields. The dielectric
contribution ($\Delta \varepsilon$, strength of the mode) and the
eigenfrequency ($\nu_0$) of the phonon are shown in the main
panels. The transition is manifested by a significant softening of
the phonon frequency and by a $60 \%$ increase of the mode
strength on increasing magnetic field. The characteristic fields and the width of the
transition closely coincide with the corresponding changes of the
dielectric constant at lower frequency, which is demonstrated in
the upper inset of Fig. \ref{fpar}. The difference in
$\varepsilon_1$ on increasing and decreasing fields documents the
meta-stable character of the IC-CA phase boundary. Similar
behavior of the high- and low-frequency dielectric constant
indicates that the same mechanism is operative for both effects,
namely that at the IC-CA transition electromagnons are suppressed
by the external magnetic field \cite{nphys}.

\begin{figure}[]
\includegraphics[width=7.5cm,clip]{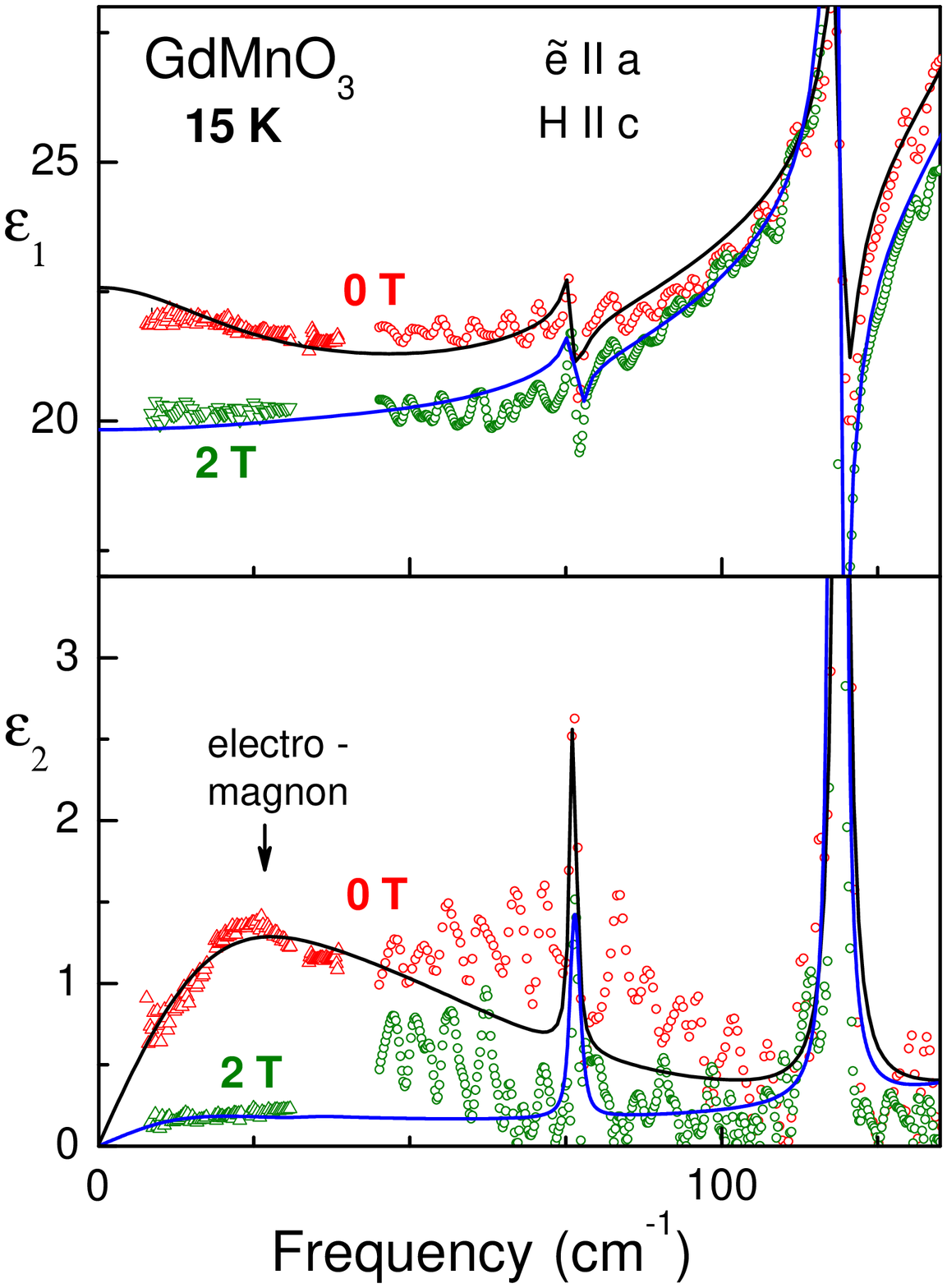}
\caption{(color online) Spectra of the a-axis dielectric
permittivity of GdMnO$_3$ in the frequency range between
electromagnon at 25 cm$^{-1}$ and phonon at 119 cm$^{-1}$ at  0\ T
and 2\ T. Upper panel: real part of the dielectric permittivity,
lower panel: imaginary part. Triangles represent the results from
the Terahertz transmittance. Circles - spectra obtained via
Kramers-Kronig analysis of the reflectance. Solid line corresponds
to the sum of Lorentzians as described in the text.} \label{feps}
\end{figure}

In order to analyze the interplay between electromagnons and
phonons the complex dielectric permittivity has been calculated
from the reflectance spectra via the Kramers-Kronig transformation
adding the Terahertz spectra at low frequencies. The results both
in IC-AFM  (15 K, 0 T) and in CA-AFM state (15 K, 2 T) are shown
in Fig. \ref{feps}. Here the data above 40 cm$^{-1}$ represent the
results of the Kramers-Kronig analysis and the data below this
frequency have been taken directly from the Terahertz
transmittance experiments. The lower panel of Fig. \ref{feps}
clearly demonstrates the over-damped almost relaxational nature of
the electromagnon and its suppression by the external magnetic
field. On the other hand, we know from Fig. \ref{fpar} that the
phonon mode at 119 cm$^{-1}$ gains considerable spectral weight on
increasing magnetic field. The substantial spectral weight which
is removed from the low-frequency range, with the dielectric
constant decreased for all frequencies below $\sim 40$ cm$^{-1}$,
is transformed into phonon intensity at 119 cm$^{-1}$. In order to
obtain the estimate of the spectral-weight transfer, the complex
dielectric permittivity has been fitted assuming the phonon
parameters as obtained from the fits of the reflectance and a
single over-damped Lorentzian representing the electromagnon with
$\nu_0 =40 $ cm$^{-1}$, $\gamma = 75$ cm$^{-1}$ and $\Delta
\varepsilon = 1.6$. Here $\gamma$ is the damping. These parameters
correspond to the maximum in $\varepsilon_2$ at $\nu_{\rm max}
\simeq 27 $ cm$^{-1}$ and to the spectral weight $S=\Delta
\varepsilon \nu_0^2 \simeq 2.5 \cdot 10^3$ cm$^{-2}$. From the
data in Fig. \ref{fpar} the
 spectral-weight increase of the phonon is obtained as $\Delta S(IC-CA)
\simeq 1.7 \cdot 10^3$ cm$^{-2}$. Taking into account strong
scattering of the data between 40 and 100 cm$^{-1}$ we conclude
rough coincidence of these values and the transformation of the
spectral weight of the electromagnon into the
 phonon mode at 119 cm$^{-1}$.

\begin{figure}[]
\includegraphics[width=8.5cm,clip]{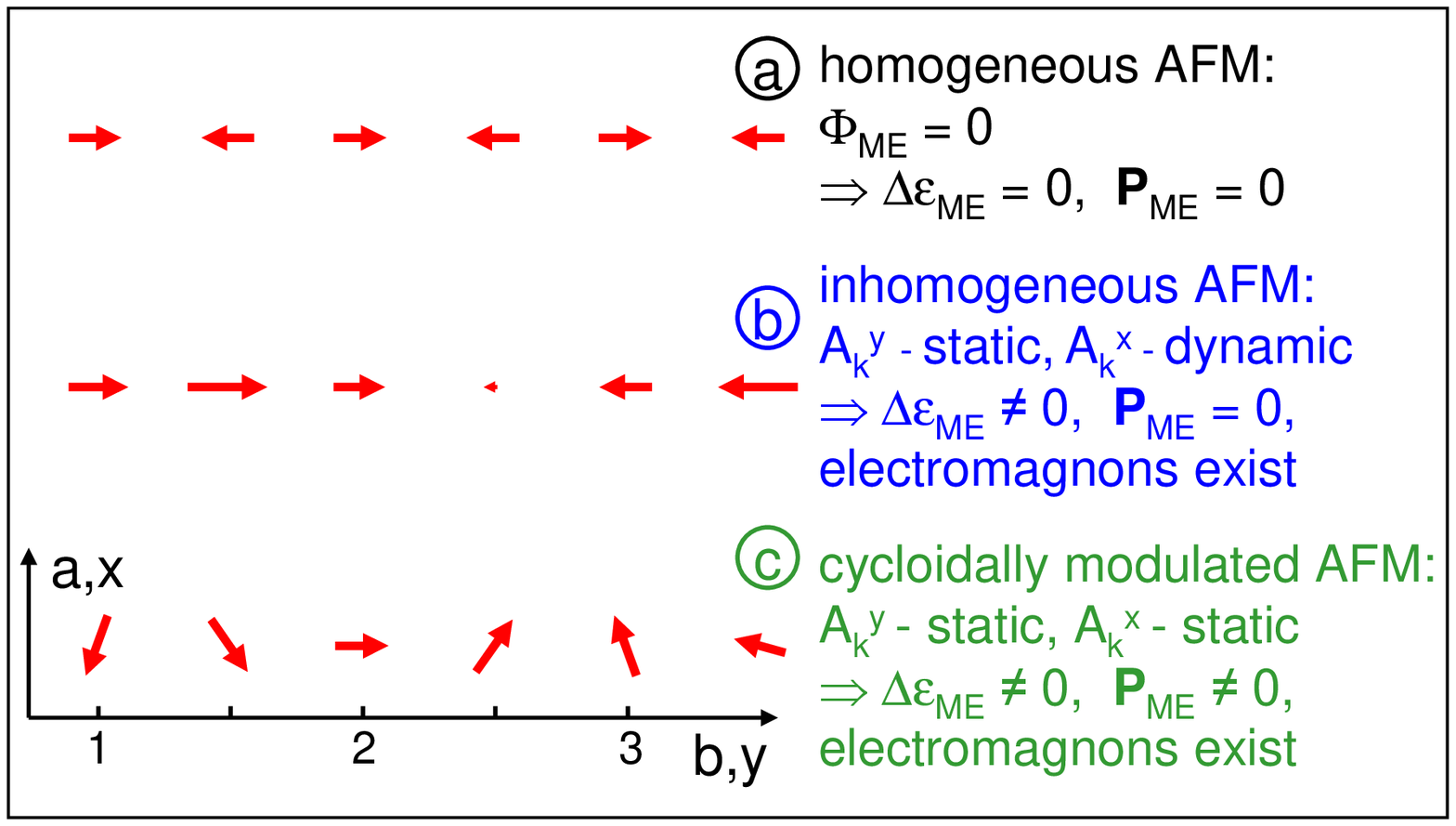}
\caption{(color online) Schematic presentation of magnetic
structures, showing different magneto-electric behavior. (a)
Homogeneously modulated structure: no static electric polarization
($\Delta P_{\rm ME} =0$), no ME contribution to the dielectric
permittivity ($\Delta \varepsilon =0$). (b) Inhomogeneously
modulated collinear structure: no electric polarization, but
nonzero contribution to the dielectric permittivity. (c)
Cycloidally modulated structure: finite electric polarization and
ME contribution to the dielectric permittivity.} \label{fschema}
\end{figure}

A remaining question is why  electromagnons are observed both in
GdMnO$_3$, in a non-ferroelectric collinear magnetic structure, as
well as in TbMnO$_3$, in a ferroelectric with helical spin
structure \cite{nphys}. In order to understand this point, Fig.
\ref{fschema} provides a closer inspection of the free-ME energy
in terms of the Fourier components of the dynamic variables ${\bf
A}_k$ related to the main AFM vector $\bf A$ of modulated spin
structure \cite{bertaut} (see also Refs. \cite{lawes,kenzelmann}):
\begin{eqnarray}\label{eqfree}
\Phi_{me} &=&  -iP_x\sum_k a_k^{xz}({\bf A}_k \times {\bf
A}_k^*)_z - \nonumber \\ & & -
iP_z\sum_k a_k^{zx}({\bf A}_k \times {\bf A}_k^*)_x  \equiv \nonumber\\
  & \equiv & - \sum_k {\bf PE}_{int}({\bf A}_k) \  ,
\end{eqnarray}
where $\bf P$ is electric polarization. The ME coefficients
$a_k^{xz, zx}$ for the nearest neighbors within the ab-plane are
determined by $a_k^{xz, zx} = 2N a^{xz, zx} \cos(2\pi kb)\sin(\pi
kb)$, where $a^{xz, zx}$ are constants, $N$ the number of Mn ions,
and $b$ the lattice constant.
This expression was derived using the crystallographic symmetry
$D_{2h}^{16}$  (Pbnm space group) and a modulated spin structure
with ${\bf k}=(0,k,0)$. We omitted weak contributions from other
AFM vectors $\bf F, C, G$ \cite{bertaut} which exist in this
structure. In space representation and continuum limit Eq.
(\ref{eqfree}) corresponds to $\Phi_{me} = - a_x P_x(A_x
\partial A_y/\partial y - A_y\partial A_x/\partial y) -
a_zP_z(A_z\partial A_y/\partial y - A_y\partial A_z/\partial y)$
\cite{nphys}, and for $a_k^{xz} = a_k^{zx}$ is reduced to
Dzyaloshinski-Moriya type interactions \cite
{katsura05,mostovoy,sergienko}.

It is clear that in a homogeneous magnetic state, like in the
CA-AFM phase (Fig. 4a), the ME free energy is zero and no
contribution to the dielectric constant, no electromagnons and no
spontaneous polarization can exist. To obtain the ME contribution
to the electric susceptibility in sinusoidal and spiral states we
consider the total free energy of the system:
\begin{eqnarray}\label{eqfree2}
    && \Phi({\bf A}_k, {\bf P}) = \qquad \nonumber \\ &=& {\textstyle \frac{1}{2}} N
     \sum_k [-J_A(k){\bf A}_k{\bf A}_k^* + K_{bc} {\bf A}_k^z{\bf A}_k^{z*} +
    K_{ba} {\bf A}_k^x{\bf A}_k^{x*}]- \nonumber \\  &&- {\bf E\cdot P} + {\bf P}^2/2\chi_E  + \Phi_{me} -TS({\bf
    A}_k) \ ,
\end{eqnarray}
where the first three terms correspond to exchange and anisotropy
energies, the forth and fifth terms represent dielectric
contributions from external electric fields $\bf E$ and the last
term is the spin entropy. By minimizing Eq. (\ref{eqfree2}) with
respect to $\bf P$ the free energy can be represented as a
function of the non-equilibrium values of $\bf A_k$. In a
sinusoidal spin structure with ${\bf A}_k = (0, {\bf A}_k^y, 0)$
and keeping only the main harmonic $k_0$ of the modulated
structure, the electric susceptibility, e.g. along the a - axis,
can be expressed as
\begin{eqnarray} \label{eqchi2}
\chi_E^x &&\approx - \left.\frac{\partial^2\Phi}{\partial E_x^2} +
\frac{\partial^2\Phi}{\partial E_x\partial A_k^x}
\frac{\partial^2\Phi}{\partial E_x\partial
A_k^{x*}}/\frac{\partial^2\Phi}{\partial A_k^x\partial A_k^{x*}}
\nonumber \right. \\ && \nonumber \\
 &&\approx  \left. \chi_E + (\chi_E a_{k0}^{xz})^2 A_{k0}^y
A_{k0}^{y*}/ K_{ba} \right. \ .
\end{eqnarray}
In the sinusoidal phase no spontaneous polarization can exist
since only one ${\bf A}_k$ component is nonzero, but the ME
contribution to electric susceptibility arises according to Eq.
(\ref{eqchi2}). It originates from an electric field-induced
rotation of the spins in the ab-plane, i.e. from ${\bf A}_k^x$
spin components. Similar contribution can also exist along c-axis.
Their relative values are determined by both, the corresponding
magnetoelectric ($a^{xz, zx}$) and magnetic anisotropy ($K_{ba,
bc}$) constants. Finally in case c) of Fig. \ref{fschema} with
helical or cycloidally modulated spins, e.g. ${\bf A}_k =(0, {\bf
A}_k^y, {\bf A}_k^z)$, a spontaneous ferroelectric polarization
along the c-axis $P_z$, finite contribution to the dielectric
constant along the a-axis $\chi_E^x \approx \chi_E + (\chi_E
a_{k0}^{xz})^2 S^2/ K_{ba}$, and electromagnons exist.


Very recently Katsura \textit{et al.} \cite{katsura} have
calculated the collective mode dynamics of helical magnets coupled
to the electric polarization. For the ac dielectric properties
their main findings are the occurrence of two modes, one of which
is derived from the phonon mode with a frequency close to the
eigenfrequency of the uncoupled phonon, and one originating from
the spin wave with a frequency proportional to $\sqrt{SJD}$, where
$S$ is the spin value, $J$ the exchange coupling and $D$ the
anisotropy. Using realistic parameters they calculate an
electromagnon frequency of 10 cm$^{-1}$, close to the experimental
observation. It also follows from their calculation that in the
electromagnetic phase the phonon eigenfrequency ($\nu_0$) is
enhanced by $\nu_p^2/2\nu_0$, where $\nu_p$ is the frequency of
the spin-plane rotation mode, which from the lower panel of Fig. 2
can now be estimated as 13 cm$^{-1}$.

In conclusion, studying the low-frequency electrodynamics of
GdMnO$_3$ with a finite magneto-electric (ME) coupling, we were
able to demonstrate i) the existence of electromagnons, ii) that
these collective modes of ME magnets contribute to the static
dielectric constant, and iii) that at the transition to a
homogeneous magnetic phase with no ME coupling the electromagnons
are wiped out and their spectral weight is transferred to an
optical phonon, which in addition reveals a slight softening of
the eigenfrequency. To conclude with an outlook for the future
applications, a material with similar effects at room temperature
will directly allow the design of a new class of magneto-optical
devices. The dielectric constant and hence the refractive index $n
\simeq \sqrt{\varepsilon_1}$ can be tuned by moderate magnetic
fields.

This work was supported by BMBF(13N6917B - EKM), by DFG (SFB 484),
and by RFBR (04-02-16592, 06-02-17514).

\end{document}